\documentclass[twocolumn,showpacs,preprintnumbers,amsmath,amssymb]{revtex4}

\usepackage{graphicx}
\usepackage{dcolumn}
\usepackage{bm}

\begin{document}

\title{Accurate spectroscopy of Sr atoms}

\author{Ir\`{e}ne Courtillot}
\author{Audrey Quessada-Vial}
\author{Anders Brusch}
\author{Dmitri Kolker}
\author{Giovanni D. Rovera}
\author{Pierre Lemonde}

\email{pierre.lemonde@obspm.fr} \affiliation{BNM-SYRTE,
Observatoire de Paris\\ 61, Avenue de l'observatoire, 75014,
Paris, France}

\date{\today}

\begin{abstract}
We report the frequency measurement with an accuracy in the
100\,kHz range of several optical transitions of atomic Sr :
$^1S_0-\,^3P_1$ at 689\,nm, $^3P_1-\,^3S_1$ at 688\,nm and
$^3P_0-\,^3S_1$ at 679\,nm. Measurements are performed with a
frequency chain based on a femtosecond laser referenced to primary
frequency standards. They allowed the indirect determination with
a 70\,kHz uncertainty of the frequency of the doubly forbidden
$5s^2~^1S_0-\,5s5p~^3P_0$ transition of $^{87}$Sr at 698\,nm and
in a second step its direct observation. Frequency measurements
are performed for $^{88}$Sr and $^{87}$Sr, allowing the
determination of $^3P_0$, $^3P_1$ and $^3S_1$ isotope shifts, as
well as the $^3S_1$ hyperfine constants.
\end{abstract}

\maketitle

\maketitle
\section{Introduction}

Precise measurements of the energy of atomic levels is of extreme importance
for modern physics. To quote a few examples, for simple atoms (H, He...) they
lead to accurate tests of quantum electrodynamics and give access to
fundamental constants such as the Rydberg
constant\,\cite{Hydrogen89,Udem97,Schwob99,Berkeland95,Pastor04}. For more
complex systems they give rise to new generations of atomic clocks with the
perspective of an improved definition of the SI second (see {\it
e.g.}\,\cite{StAndrews}), they allow stringent laboratory tests of local
position invariance\,\cite{Uzan03,Bize03,Marion03,Fischer04,Peik04} and
combined with mass ratio measurements they lead to accurate determinations of
the fine structure constant $\alpha$\,\cite{Udem99,Wicht02,Gupta02,Battesti04}.

Many of the atomic transitions of interest lie in the optical domain of the
electromagnetic spectrum. In the heroic times of optical frequency metrology,
measurements were based on phase coherent harmonic frequency chains which
required a specific complex apparatus for each transition under
consideration\,\cite{Jennings83,Clairon85,Andreae92,Nez92,Schnatz96,Bernard99}.
Very few atomic or molecular resonances have been measured directly with this
technique\,\cite{Bauch02}. A few more transitions have been indirectly measured
with decent accuracy thanks to their coincidental vicinity to one of these few
references. The vast majority of atomic transitions however are only known to
within a few hundred MHz, limited by the mechanical quality of spectrometers.
The situation has dramatically changed over the last few years with the advent
of optical frequency combs\,\cite{Bauch02,Cundiff03}. Very compact and simple
apparatus are now available which in principle allow the phase coherent
comparison of any frequencies ranging from the GHz to the PHz domain.

\begin{figure}
\begin{center}
\includegraphics[height = \columnwidth, angle = -90]{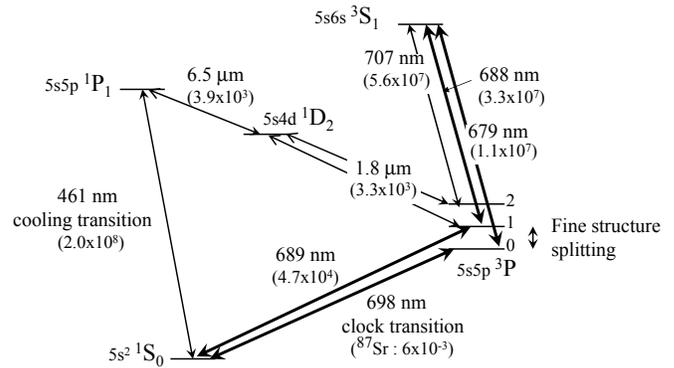}
\end{center}
\caption{Energy diagram of Sr with wavelength and decay rate
(s$^{-1}$) of the transitions involved in the experiment. Two
isotopes have been studied, $^{88}$Sr and $^{87}$Sr of natural
abundance 82\,\% and 7\,\% respectively. $^{87}$Sr possesses a
hyperfine structure ($I=9/2$) which is not shown here for clarity.
The 698\,nm transition is weakly allowed for $^{87}$Sr by
hyperfine coupling.}\label{fig:levels}
\end{figure}

We describe here the extensive use of this technique for the
spectroscopy of Sr atoms. We report the frequency measurement with
an accuracy in the hundred kHz range of the $^1S_0-\,^3P_1$,
$^3P_1-\,^3S_1$ and $^3P_0-\,^3S_1$ transitions at 689, 688 and
679 nm respectively (see Fig.\,\ref{fig:levels}). The measurements
are performed with both 88 and 87 isotopes. For the latter,
hyperfine components have been measured giving access to the
hyperfine parameters of the $^3S_1$ state and to the isotope
shifts of $^3P_0$, $^3P_1$ and $^3S_1$. The interest of these
measurements is illustrated by the first observation of the
$^1S_0-\,^3P_0$ transition of $^{87}$Sr which we have reported
in\,\cite{Courtillot03PRA}. This was made possible by the indirect
determination of the $^3P_0$ state energy given by the
measurements detailed here. The direct frequency measurement with
a 15\,kHz accuracy of this transition in turn nicely confirms the
indirect determination. It also opens the way to a new generation
of optical frequency standards with atoms confined in a dipole
trap\,\cite{KatoPal03}.

\section{Experimental setup\label{sec:setup}}

\begin{figure}
\begin{center}
\includegraphics[width = \columnwidth]{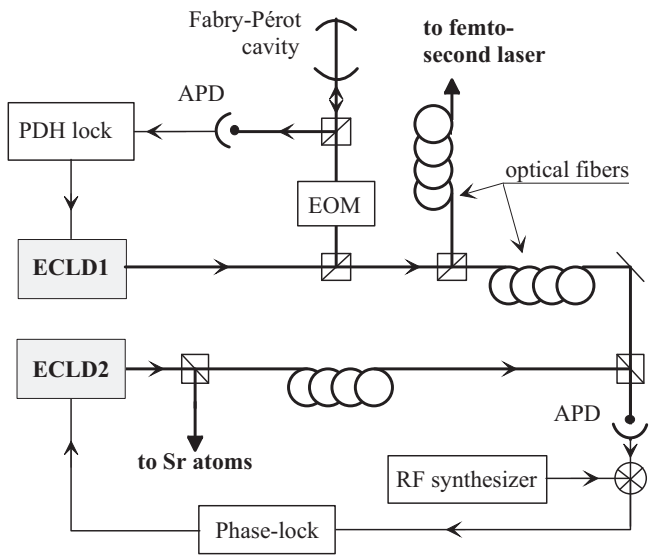}
\end{center}
\caption{Experimental set up used for frequency measurements. A
first extended cavity laser diode (ECLD1) is locked to a high
finesse Fabry-P\'{e}rot cavity. 100\,$\mu$W of its output power
are sent to a frequency chain based on a femtosecond laser for
absolute frequency measurement. A second laser (ECLD2),
offset-phase locked to ECDL1, is used to probe the strontium
atoms. Noise and frequency offsets introduced by optical fibers
are negligible in this experiment. EOM: electro-optical
modulator.} \label{fig:exp}
\end{figure}

All frequency measurements were performed with the experimental
set up sketched in Fig.\,\ref{fig:exp}. An extended cavity laser
diode (ECLD1) is locked to a high finesse cavity using the
Pound-Drever-Hall method\,\cite{Drever83} with performances
reported in Ref.\,\cite{Quessada03}. This laser has a linewidth of
35\,Hz. Its absolute frequency is continuously measured vs
BNM-SYRTE primary frequency standards with a scheme based on a
self referenced femtosecond Ti:Sapph
laser\,\cite{Bauch02,Cundiff03}. The fractional resolution of this
measurement is typically $3\times 10^{-13}$ for a one second
averaging time (corresponding to 100\,Hz in the region of the
spectrum involved here). ECLD1 is used as a frequency reference
for a second laser, ECLD2, that probes the strontium atoms. ECLD2
is offset-phase locked to ECLD1: The beat-note between both lasers
is detected by an avalanche photodiode (APD) and mixed with the
output of a radio-frequency synthesizer to generate the offset
phase lock error signal. The bandwidth of the servo control is
2\,MHz. By actuating the RF synthesizer, the light of ECLD2 which
is sent to the atoms can be continuously tuned to any frequency
between two modes of the cavity of free spectral range 1.5\,GHz.
This point was of importance for the first detection of the
transitions since their frequencies were not accurately known. To
measure all the atomic resonances, two sets of lasers are used.
The first set can be tuned from 675 to 685\,nm, the second one
from 685 to 698\,nm.

Except for the $^1S_0-\,^3P_1$ transition, the frequency
measurements are made by probing a sample of cold atoms collected
in a magneto-optical trap (MOT). An atomic beam is decelerated in
a Zeeman slower and captured at the crossing-point of three
retro-reflected beams tuned to the red of the $^1S_0- \,^1P_1$
transition at 461\,nm. The setup is detailed in
Ref.\,\cite{Courtillot03OL}. For these experiments the MOT is
typically loaded with  $10^7$ atoms of $^{88}$Sr or $10^6$ atoms
of $^{87}$Sr. At steady state, the atomic cloud has a $1/e^2$
diameter of about $2$\,mm and a temperature of 2\,mK.

\section{frequency measurements of the $^1S_0-\,^3P_1$ transition}

\begin{figure}
\begin{center}
\includegraphics[width=\columnwidth]{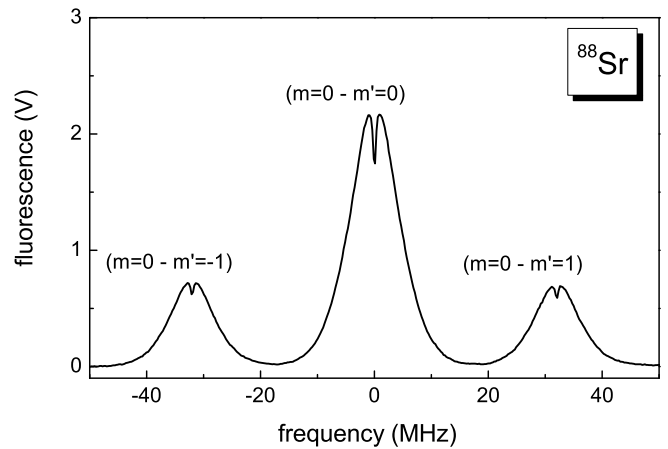} 
\end{center}
\caption{$^1S_0-\,^3P_1$ resonance of $^{88}$Sr. The Lamb dip at
the center of the Doppler profile is broadened by saturation to
550\,kHz. All three Zeeman components are visible on this scan.}
\label{fig:res_88_689}
\end{figure}

The $^1S_0-\,^3P_1$ transition at 689\,nm is easy to detect since
it involves the ground state. We use saturated fluorescence in an
atomic beam which design was previously described in
Ref.\,\cite{Courtillot03OL}. This atomic beam is collimated to a
diameter of 4\,mm and a divergence of 12 mrad (half-width at half
maximum). The atomic flux is about $10^{12}$ at.s$^{-1}$ at an
average velocity close to 500\,m.s$^{-1}$. Atoms are probed by a
retro-reflected laser beam orthogonal to the atomic beam. A
typical spectrum recorded for $^{88}$Sr is shown in
Fig.\ref{fig:res_88_689}. For frequency measurements, the laser is
locked to the Lamb dip at the center of the Doppler profile.

The Zeeman sub-levels are split by a static magnetic field
parallel to the atomic beam. The scan in
Fig.\,\ref{fig:res_88_689} is recorded with a laser polarization
forming a non-zero angle with the static magnetic field of
1.5\,mT. All three Zeeman components of $^{88}$Sr are then
visible. Frequency measurements are performed on the
{\it{m}}\,=\,0\,-\,{\it{m'}}\,=\,0 transition, insensitive to the
first order Zeeman effect.

\begin{figure}
\begin{center}
\includegraphics[width=\columnwidth]{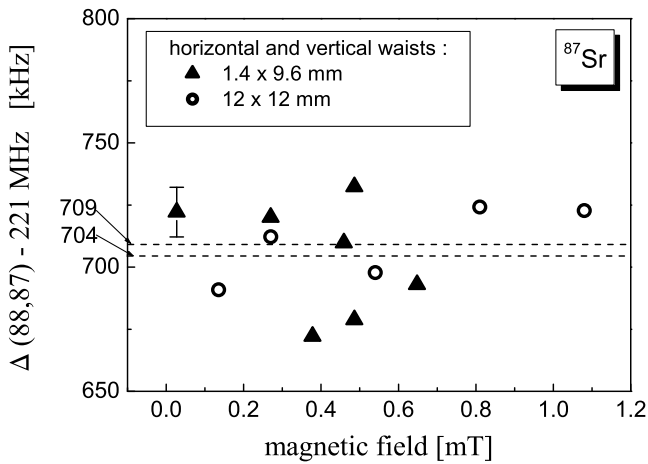}
\end{center}
\caption{\small{Frequency measurements of the
$^1S_0,\,F=9/2\,-\,^3P_1,\,F'=9/2$ transition of $^{87}$Sr
referenced to the frequency of $^1S_0-\,^3P_1$ of $^{88}$Sr as a
function of the magnetic field. The mean values of both sets of
measurements are indicated by dashed lines. The measurements are
performed with a laser power of 14\,mW. The statistical
uncertainty of each point is 10\,kHz. The "horizontal" waist is
parallel to the atomic beam.}}\label{fig:87Sr689}
\end{figure}

For $^{87}$Sr, which possesses a nuclear spin $I=9/2$, all
possible transitions depend on the magnetic field to first order.
The frequency shift of state $|^3P_1, F=9/2,m\rangle$ is $m\times
0.8$\ MHz/mT. However, one expects all Zeeman sub-levels of the
ground state to be equally populated in the thermal beam. In
addition, since the magnetic field orientation is orthogonal to
the probe laser propagation axis, the laser polarisation has equal
$\sigma^+$ and $\sigma^-$ components. One then expects the
sub-Doppler structure of the resonance to be fully symmetric.
Insensitivity of the frequency measurements to the magnetic field
is confirmed experimentally (Fig.\,\ref{fig:87Sr689}).

\begin{figure}
\begin{center}
\includegraphics[width=\columnwidth]{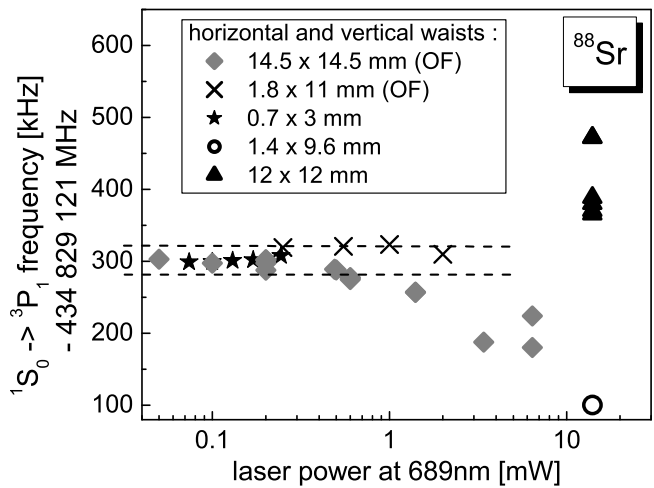}
\end{center}
\caption{\small{Frequency measurements of $^1S_0-\,^3P_1$
transition for $^{88}$Sr, with various laser beam geometries. The
horizontal waist is parallel to the atomic beam. (OF) : an optical
fiber is used as a spatial filter of the laser mode. The
statistical error of all the measurements is less than 1\,kHz.
}}\label{fig:doppler689}
\end{figure}

The main effect in the accuracy budget is the residual first order
Doppler frequency shift. It is minimized by a careful alignement
of the incident and retroreflected beams. First, the incident beam
is collimated with a shear plate interferometer. It is aligned
perpendicular to the atomic beam by equalizing to better than
20\,kHz the central frequency of the Doppler profile in single
pass operation and the Lamb dip frequency. Second, the laser beam
is retro-reflected by a cat eye allowing for a fine cancellation
of the angle between the two counterpropagating laser beams. The
cat eye is designed with a 15\,cm focal length lens and a plane
mirror located at the beam waist. It is adjusted by measuring the
Lamb dip frequency as a function of the lens position: for the
optimal lens-mirror distance, the frequency is independent on the
transverse position of the lens. Experimentally we observe
frequency shifts lower than 1\,kHz for transverse displacements of
the lens up to $\pm 1$mm. First order Doppler effect also arises
at high laser power from wavefront distortion due to imperfect
beam collimation and
aberrations\,\cite{Hall76,Ishikawa94,Lemonde98}. We put an upper
bound to this effect by performing frequency measurements as a
function of the probe laser power with several laser beam
geometries (leading to different wave fronts). The results of this
study which was conducted for $^{88}$Sr are summarized in
Fig.\,\ref{fig:doppler689}. At low laser power the measurements
performed in all geometries tend towards the same frequency to
within $\pm$20\,kHz, while at high power a dispersion of hundreds
of kHz is observed.  Our final value for $^{88}$Sr is the mean
value of the measurements performed at a power below 1\,mW, to
which we attribute a conservative error bar of 20\,kHz : 434 829
121 300 (20)\,kHz. It is in good agreement with the measurement
performed simultaneously by Ferrari {\it et al} \cite{Ferrari03}
(434 829 121 311 (10)\,kHz).

For $^{87}$Sr frequency measurements have not been done at these
low laser powers. Due to its small natural abundance (7 \%) and to
the Zeeman degeneracy of the ground state the sub-Doppler
structure is hardly detectable in the absence of saturation
broadening. The atomic frequency is obtained by comparison to
$^{88}$Sr: at high laser power, we alternate measurements of the
$^{87}$Sr and $^{88}$Sr resonances. We have checked that the
difference between both atomic frequencies does not depend on the
laser beam geometry to within 50\,kHz (Fig.\,\ref{fig:87Sr689}).
Only the hyperfine component of $^{87}$Sr with $F'=9/2$ has been
measured with this technique thanks to its small frequency
difference with the $^{88}$Sr transition. Since the hyperfine
structure of the $^3P_1$ state of $^{87}$Sr is very well
known\,\cite{Putlitz63} this particular measurement allows the
determination of the two other hyperfine components (see table
\,\ref{tab:freq689}). One can also extract the isotope shift of
state $^3P_1$, $\Delta_{87,88}\,[^3P_1] = 62\,150\,(70)$\,kHz,
with an accuracy improvement by more than one order of
magnitude\,\cite{Buchinger85,Celikov95}.

\begin{table}
\begin{center}
\begin{tabular}{c|c|lc}
\hline  \hline
\multicolumn{2}{c}{$^1S_0-\,^3P_1$} &  \multicolumn{2}{c}{Frequency (kHz)}\\
  \hline
$^{88}$Sr &J=0\,-\,J'=1& 434 829 121 300 (20)&(a)\\
\hline 
 & F=9/2\,-\,F'=7/2 & 434 830 473 270 (55)&(b)\\
\cline{2-4}
$^{87}$Sr& F=9/2\,-\,F'=9/2 & 434 829 343 010 (50)&(a)\\
\cline{2-4}
& F=9/2\,-\,F'=11/2 & 434 827 879 860 (55)&(b)\\
\hline  \hline
\end{tabular}
\end{center}
\caption{\small{$^1S_0-\,^3P_1$ atomic frequencies. (a) : Direct
measurements. (b) : Values computed from the measurements reported
here and from the $^3P_1$ hyperfine constants given in
Ref.\,\cite{Putlitz63}.}}\label{tab:freq689}
\end{table}

\section{frequency measurements of the $^3P_1-\,^3S_1$ transition}

Measurements of the $^3P_1-\,^3S_1$ transition at 688 nm are
performed with the cold atoms in the MOT in which the $^3P_1$
state is slightly populated. Indeed, while cycling on the
$^1S_0-\,^1P_1$ transition, atoms occasionally decay to $^1D_2$ by
spontaneous emission (Fig.\,\ref{fig:levels}). This state has a
lifetime of 0.3\,ms and two main decay channels to $^3P_1$ and
$^3P_2$, with respective branching ratios 67\% and 33\%
\cite{Bauschlicher85}. Atoms reaching $^3P_1$ decay back to the
ground state fast enough that they are kept in the trap:
altogether this radiative cascade has a duration of a few hundred
microseconds during which atoms travel a distance smaller than the
MOT size. On the other hand atoms in the metastable $^3P_2$ state
are lost for our experiment which limits the lifetime of the MOT
to 30 to 50\,ms, depending on the trapping laser saturation.

\begin{figure}
\begin{center}
\includegraphics[width=\columnwidth]{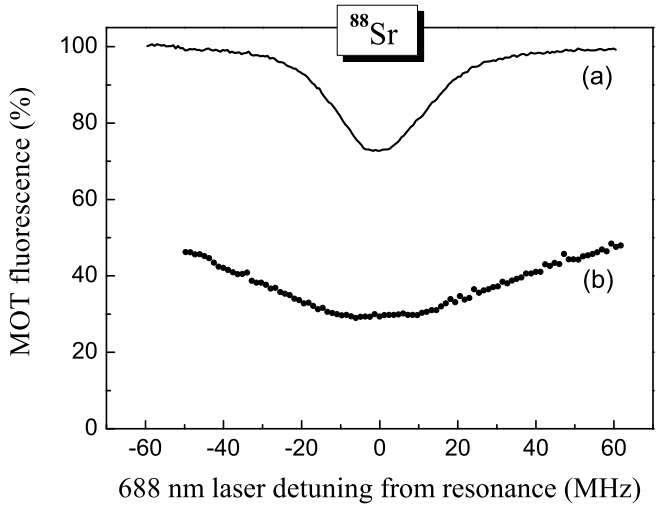}
\end{center}
\caption{Relative MOT fluorescence signals obtained when a 688\,nm
laser is swept around the $^3P_1-\,^3S_1$ transition of $^{88}$Sr.
Measurements performed with a 0.9\,mm beam waist radius and laser
intensities at the center of the beam of (a) $I_{688} =
0.2$\,mW.cm$^{-2}$ and (b) $I_{688} = 51$\,mW.cm$^{-2}$. The MOT
has a $1/e^2$ diameter of 2 mm.} \label{fig:688}
\end{figure}

The $^3P_1-\,^3S_1$ transition is detected by modifying this
radiative process. If a laser resonant to one of the
$^3P_1-\,^3S_1$ transitions is added to the trap, atoms in the
corresponding $^3P_1$ state are pumped to $^3P_2$ and $^3P_0$
metastable states. They escape the trap instead of decaying back
to the ground state. This additional loss mecanism decreases the
trapped atom number. Figure\,\ref{fig:688} shows the relative MOT
fluorescence (at 461\,nm) as a function of the 688\,nm laser
detuning from the $^3P_1-\,^3S_1$ resonance of $^{88}$Sr. Similar
resonances are obtained with all hyperfine components of
$^{87}$Sr. It should be noted that although the relative
population of the $^3P_1$ state is lower than $10^{-3}$, the
signal is very large indeed. The decrease of the MOT blue
fluorescence can reach $40$\% for $^{87}$Sr and $70$\% for
$^{88}$Sr at sufficiently high laser intensity.

Atoms are probed in a standing wave configuration by
retroreflecting the 688\,nm laser beam which shines the trapped
atoms. The two counterpropagating beams are aligned to better than
$1$\,mrad. The residual Doppler effect due to a possible asymmetry
of the velocity distribution of the trapped atoms is well below
1\,kHz. We have studied the contrast and width of the resonances
as a function of the 688\,nm laser intensity for probe beam waists
ranging from 0.75 to 6 mm. Theoretical values can be simply
derived from the rate equations of the MOT dynamics if one assumes
that the capture process is not affected by the probe beam. We
observe a strong discrepancy between experimental data and the
model. The experimental resonance width is about ten times smaller
than expected. Furthermore, the required laser intensity for which
the contrast is half its maximal value is two orders of magnitude
higher than predicted.

This shows that the effective laser intensity is much smaller than
its value in the model. A significant fraction of the atoms
contributing to the signal are presumably pumped at the edge of
the probe beam. They escape the trap before reaching the central
region of the trap. An accurate model accounting for this effect
would require a good knowledge of the spatial and velocity
distributions of the atoms during capture which is a hard task for
a MOT loaded with a Zeeman slower.

The main consequence of numerous atoms being excited away from the
trap center is a high sensitivity of the frequency measurements to
the Zeeman effect induced by the MOT magnetic field. The magnetic
field gradient is 1.8\,mT/cm and its effect on the measured
frequency is difficult to evaluate because the Zeeman structure is
not resolved in the experiment. The actual resonance shape and
central frequency then depends on the laser polarisation and on
hardly known parameters such as the position of the MOT with
respect to the zero of the magnetic field, the population of the
various Zeeman sub-levels and the atom spatial distribution in the
capture process. Our uncertainties are deduced from experimental
observations as described in the next paragraphs. Frequencies of
the atomic resonances at 688\,nm could be obtained with an
accuracy of 300\,kHz and 500\,kHz for $^{87}$Sr and $^{88}$Sr
respectively. $^{88}$Sr is more sensitive to the Zeeman effect due
to its higher Land\'{e} factors (see Tab.\,\ref{tab:annexeLande}).

\begin{figure}
\begin{center}
\includegraphics[width=\columnwidth]{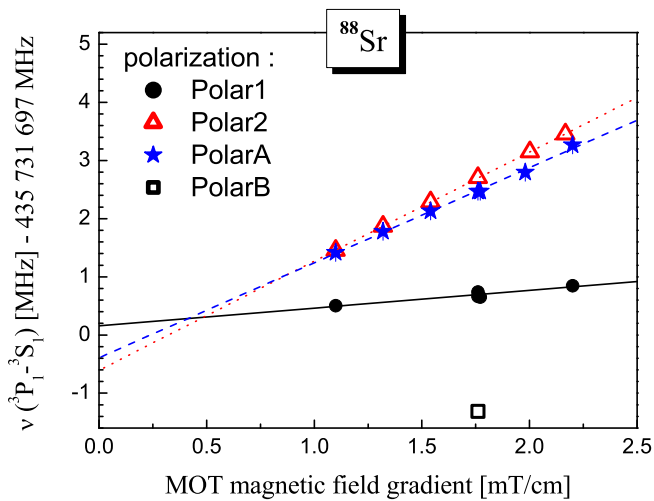}
\caption{\small{Frequency shifts due to first order Zeeman effect
induced by the MOT magnetic field. Measurements shown here are
performed for the $^3P_1-\,^3S_1$ transition of $^{88}$Sr with
various probe laser polarizations. {\sf{Polar1}} ({\sf{PolarA}})
is perpendicular to {\sf{Polar2}} ({\sf{PolarB}}). The laser
intensity at the center of the beam is 0.7\,mW.cm$^{-2}$. Typical
resolution of the frequency measurements is
20\,kHz.}}\label{fig:ImotBoson688}
\end{center}
\end{figure}

\begin{figure}
\begin{center}
\includegraphics[width=\columnwidth]{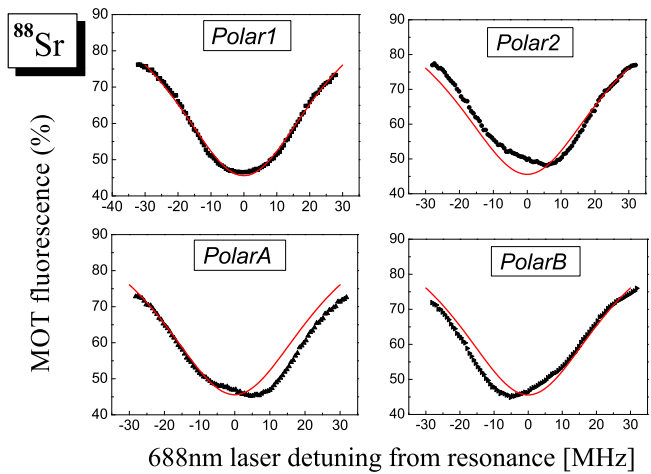}
\caption{\small{$^3P_1-\,^3S_1$ resonances of $^{88}$Sr observed
for various laser polarizations. Continuous lines in all plots
represent a fit of the resonance detected with {\sf{Polar1}} (see
text).}}\label{fig:asymetrieBoson688}
\end{center}
\end{figure}

Fig.\,\ref{fig:ImotBoson688} shows the measured frequency of the
$^{88}$Sr resonance as a function of the MOT magnetic field
gradient for various probe polarisations. For each measurement the
probe laser is locked to the resonance with a square frequency
modulation, a technique which is sensitive to an eventual
asymmetry of the resonance. Although measurements with different
polarizations come closer and closer while decreasing the field
gradient, extrapolation to zero is not perfect. We attribute this
behaviour to the asymmetry of the resonances for polarizations
other than the one labelled {\sf{Polar1}}
(Fig.\,\ref{fig:asymetrieBoson688}). For this polarization, the
resonance is nearly symmetric and the sensitivity to the Zeeman
effect is minimal (300\,kHz/[mT/cm]). For both $^{87}$Sr and
$^{88}$Sr, we take as a final frequency the extrapolated value at
zero field of the measurements performed with {\sf{Polar1}}. The
final error bar reflects the dispersion of the various zero field
extrapolations. For $^{87}$Sr, we observe a smaller dependence of
the frequency on the MOT field gradient. Among all the $^{87}$Sr
transitions at 688\,nm, the highest dependency was observed for
$^3P_1,F=7/2-\,^3S_1, F'=9/2$: 150\,kHz/[mT/cm] in polarization
{\sf{Polar1}}.

\begin{figure}
\begin{center}
\includegraphics[width=\columnwidth]{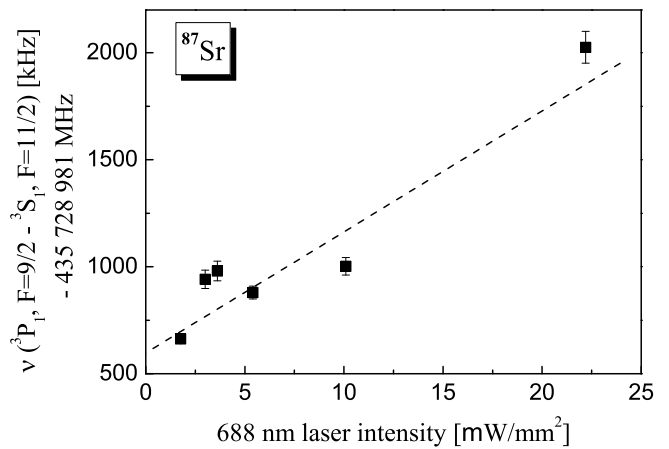}
\caption{\small{Frequency shifts measured as a function of the
laser intensity. Measurements performed on the $^3P_1,
F=9/2-\,^3S_1,F=11/2$ transition of $^{87}$Sr with a laser waist
radius of 0.9\,mm and a MOT field gradient of
1.8\,mT/cm.}}\label{fig:effetP688}
\end{center}
\end{figure}

The effect of the magnetic field gradient can also be seen by
varying the 688\,nm laser intensity (Fig.\,\ref{fig:effetP688}).
At increasing laser powers atoms further and further away from the
trap center (where the magnetic field is higher) contribute to the
signal. In order to minimize this effect, the final frequency is
extrapolated from measurements done at low laser intensity only,
about 1\,mW.cm$^{-2}$.

\begin{table}
\begin{center}
\begin{tabular}{c|c|l}
\hline  \hline
 \multicolumn{2}{c|}{$^3P_1-\,^3S_1$}  &  \multicolumn{1}{c}{Frequency (kHz)}\\
 \hline
    $^{88}$Sr & $J= 9/2, J'=9/2$ &  435 731 697 200 (500) \\
\hline
& $F=7/2 - F'=7/2$  &   435 733 271 100 (600) \\ 
\cline{2-3}
&  $F=7/2 - F'=9/2$ &  435 730 832 300 (300)\\
\cline{2-3}
  &$F=9/2 - F'=7/2$ &   435 734 401 750 (300)  \\
\cline{2-3}
 $^{87}$Sr &$F=9/2 - F'=9/2$  & 435 731 962 700 (300) \\
\cline{2-3}
  &$F=9/2 - F'=11/2$   &  435 728 981 600 (300)\\
\cline{2-3}
  &$F=11/2 - F'=9/2$  & 435 733 425 800 (300)\\
\cline{2-3}
 & $F=11/2 - F'=11/2$  &   435 730 444 900 (300)\\
\hline  \hline
\end{tabular}
\end{center} \caption{\small{$^3P_1 -\,^3S_1$ frequency measurements.}}\label{tab:freq688}
\end{table}

\begin{table}
\begin{center}
\begin{tabular}{c|c|c|c}
\hline \hline & \multicolumn{2}{c|}{This work}& Ref.\,\cite{Putlitz63}\\
\raisebox{1.5ex}[0cm][0cm]{$F_1 - F_2$}
 & via $|^3S_1, F'\rangle$&  kHz& kHz\\\hline
     &   7/2 & 1 130 650 (800)& \\
\raisebox{1.5ex}[0cm][0cm]{7/2 - 9/2} & 9/2& 1 130 400 (600)& \raisebox{1.5ex}[0cm][0cm]{1 130 260 (20)} \\
\hline
    &     9/2 & 1 463 100 (600)& \\
\raisebox{1.5ex}[0cm][0cm]{9/2 - 11/2} &11/2 &1 463 300 (600) & \raisebox{1.5ex}[0cm][0cm]{1 463 150 (20)} \\
\hline
    7/2 - 11/2 &  9/2 & 2 593 500 (600)&  2 593 410 (20) \\
\hline\hline
\end{tabular}
\caption{\small{For $^{87}$Sr : Frequency splitting between hyperfine levels
$|^3P_1,\,F_1\rangle$ and $|^3P_1,\,F_2\rangle$, computed with the measurements
at 688\,nm involving the hyperfine level $|^3S_1,\,F'\rangle$. Uncertainties of
individual measurements are conservatively added linearly: although the Zeeman
effects are expected to be quite different for different transitions (because
of the Land{\'e} factors) they are certainly not independent on each other.
}}\label{tab:hyperfine_3P1}
\end{center}
\end{table}

In Table\,\ref{tab:freq688} are reported the frequencies obtained
for all transitions at 688\,nm\footnote{The particular 600\,kHz
uncertainty attributed to the $^3P_1,F=7/2-\,^3S_1, F'=9/2$
transition is due to the fact that this resonance is less than
$100\,$MHz away from the nearest mode of the reference Fabry-Perot
cavity. In this case, the offset phase lock loop is disturbed and
the spectrum of the probe laser is slightly asymmetric.}. These
measurements can be partly cross-checked by extracting the
accurately known hyperfine splitting of the $^3P_1$ state of
$^{87}$Sr\,\cite{Putlitz63}. Values are reported in
Table\,\ref{tab:hyperfine_3P1}. The agreement between values
deduced from our measurements via different $^3S_1,F'$ states on
the one side and between our values and those of
Ref.\cite{Putlitz63} on the other side, strongly assesses the
conservativeness of our uncertainties.

Frequency measurements of $^1S_0 -\,^3P_1$ and $^3P_1 -\,^3S_1$
transitions give the isotope shift, $\Delta_{87,88}\,[^3S_1]$
(referenced to the ground state), and the hyperfine constants $A$
and $B$ of the $^3S_1$ state:
\begin{eqnarray}
\nonumber && \Delta_{87,88}\,[^3S_1] = 54.9\;(3)\,\mathrm{MHz}\\
\nonumber &&A\,[^3S_1]= -542.0\;(1)\,\mathrm{MHz}\\
\nonumber &&B\,[^3S_1]= -0.1\;(5)\,\mathrm{MHz}
\end{eqnarray}
To our knowledge, these parameters had never been measured before
this work. The value of the electric quadrupole hyperfine constant
$B$ is compatible with zero as expected when both valence
electrons are in $s$ configurations ($5s\,6s$) \cite{Sobelman79}.

\section{frequency measurements of the $^3P_0-\,^3S_1$ transition}

A direct detection of the $^3P_0-\,^3S_1$ transition by the
previous technique is not possible. Unlike $^3P_1$, $^3P_0$ is not
populated in the MOT. We have performed the measurement in two
steps. First, we induce a light shift on the $^3P_1-\,^3S_1$
transition with a 679\,nm laser close to the $^3P_0-\,^3S_1$
resonance (see Fig.\,\ref{fig:levels}). This allows the
determination of the frequency of the atomic transition at 679\,nm
to better than 1\,MHz. In a second step, we use coherent
population trapping resonances in the $\{^3P_0,\,^3P_1,\, ^3S_1\}$
$\Lambda$ system for a direct measurement of the $^3P_0-\,^3P_1$
fine structure splitting. This second step which could only be
applied to $^{87}$Sr, leads to an uncertainty of 50\,kHz on the
fine structure measurement.

For both steps, the two sets of lasers as described in section
\ref{sec:setup} are simultaneously used. At the MOT location the
beam waist radii of the probe lasers (ECLD2 in
Fig.\,\ref{fig:exp}) are $w_{679}$=1.3\,mm and $w_{688}$=0.9\,mm
respectively. 679\,nm and 688\,nm reference lasers (ECLD1) are
locked to two modes of the same high finesse cavity. From absolute
frequency measurements of one of the modes, the frequency of the
other one can be inferred to better than 1\,kHz. It has been
checked with the femtosecond frequency chain by alternating
frequency measurements of the 679\,nm and 688\,nm lasers or by
directly measuring their frequency difference.

\begin{figure}
\begin{center}
\includegraphics[width=0.49\columnwidth]{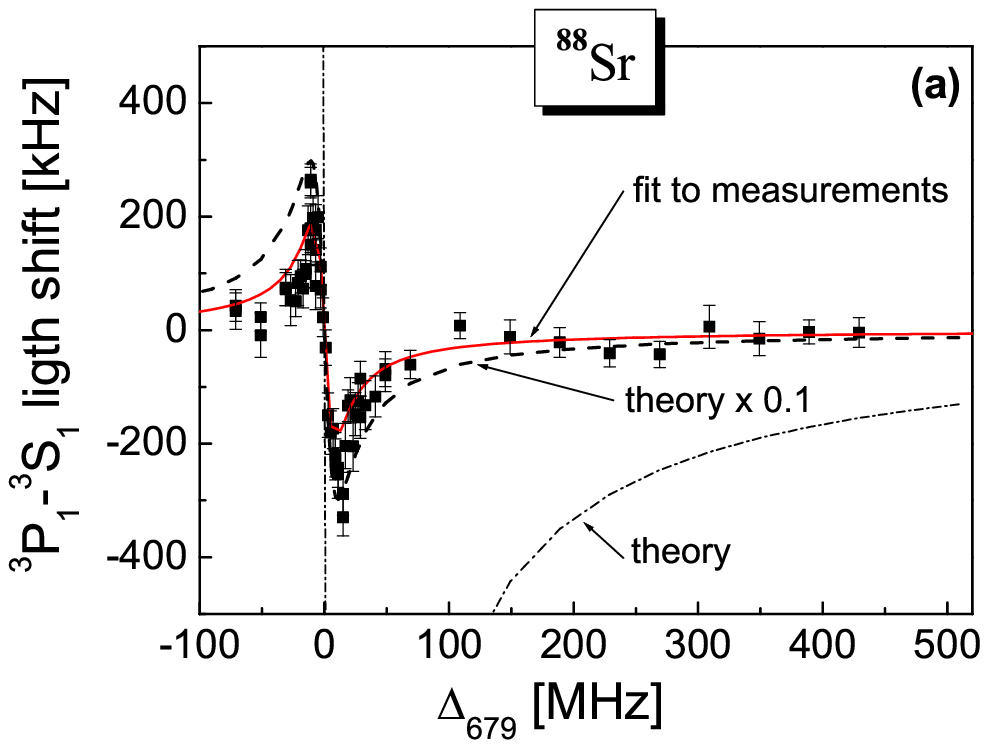}
\includegraphics[width=0.49\columnwidth]{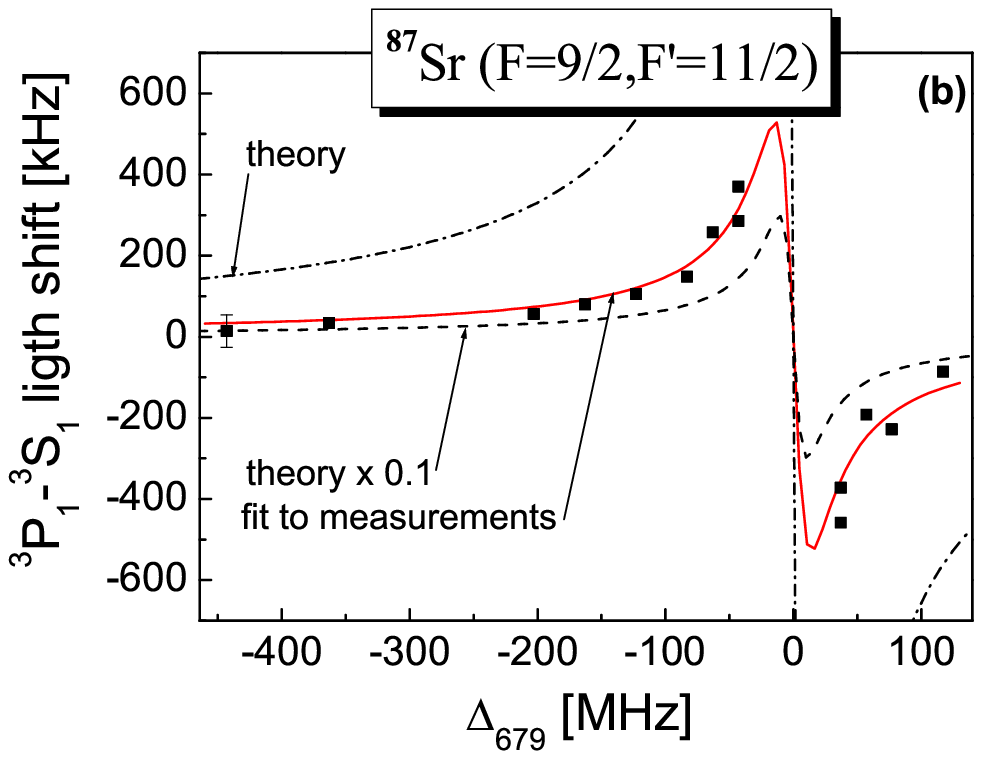}
\end{center}
\caption{\small{Light shift of the $^3P_1-\,^3S_1$ transition
induced by the 679\,nm laser as a function of $\Delta_{679}$, the
frequency detuning from $^3P_0-\,^3S_1$ resonance. The zero on the
vertical axis corresponds to the $^3P_1-\,^3S_1$ frequency
measured in the absence of  the 679\,nm laser. Measurements are
performed with mean intensities $I_{679}$=180\,mW.cm$^{-2}$ and
$I_{688}$=0.2\,mW.cm$^{-2}$. Dashed lines represent the
theoretical light shift for an atom modelled as a two-level system
with a natural linewidth $\Gamma^*/2\pi$=15.9\,MHz and a Rabi
frequency $\Omega_{679}/2\pi$\,=\,16.3\,MHz.
}}\label{fig:lightShift}
\end{figure}

Light shift measurements are plotted in
Fig.\,\ref{fig:lightShift}. The 679\,nm laser induces a frequency
dependent light shift of $^3S_1$ which is deduced from
measurements of the $^3P_1-\,^3S_1$ transition with the 688\,nm
laser. The $^3P_0-\,^3S_1$ atomic frequency corresponds to the
center of symmetry of the light shift curve. For the signal to be
as large as possible the total available power at 679\,nm is used.
A 2.4\,mW laser beam shines the trapped atoms in a standing wave
configuration. In order to minimize the number of atoms being
excited far from the MOT center the 688\,nm laser has a low
intensity, typically $I_{688}$=0.2\,mW.cm$^{-2}$. It should be
noted that the measured light shift is about 10 times smaller than
the value derived from a simple two-level atom model assuming a
transition natural width
$\Gamma^*=\Gamma_{679}+\Gamma_{688}+\Gamma_{707}= 2\pi\times
15.9\,\mathrm{MHz}$ and a Rabi frequency $\Omega/2\pi=16.3\,$MHz.

For $^{88}$Sr (Fig.\,\ref{fig:lightShift}.a) the frequency of the
resonance at 679\,nm is obtained with a statistical uncertainty of
400\,kHz. We add a 500\,kHz uncertainty due to the Zeeman effect
and end up with:
\[^{88}\mathrm{Sr}:\quad^3P_0\rightarrow\,^3S_1\quad \nu=441\,332\,751.3\,(0.7)\,\mathrm{MHz}.
\]

A more accurate measurement can be performed with coherent
population trapping\,\cite{Alzetta76}. Consider the
$\{^3P_0,\,^3P_1,\, ^3S_1\}$ $\Lambda$ system coupled with the
lasers at 679\,nm and 688\,nm. When the frequency difference
between both lasers matches the $^3P_0-\,^3P_1$ fine structure
splitting, there exists a linear superposition $|\psi_{NC}\rangle$
of the two $^3P_{0,1}$ states which is not coupled to $^3S_1$. Let
$|\psi_{C}\rangle$ be the superposition of $^3P_0$ and $^3P_1$
which is orthogonal to $|\psi_{NC}\rangle$. Atoms decaying from
$^1D_2$ are projected either onto $|\psi_{NC}\rangle$ or onto
$|\psi_{C}\rangle$. Atoms in $|\psi_{C}\rangle$ are essentially
pumped to $^3P_ 2$ and escape the trap (see
Fig.\,\ref{fig:levels}) while atoms projected onto the dark state
$|\psi_{NC}\rangle$ end up in the atomic ground state because of
the instability of $^3P_1$. These atoms recover the trapping
process.

\begin{figure}
\begin{center}
\includegraphics[width=\columnwidth]{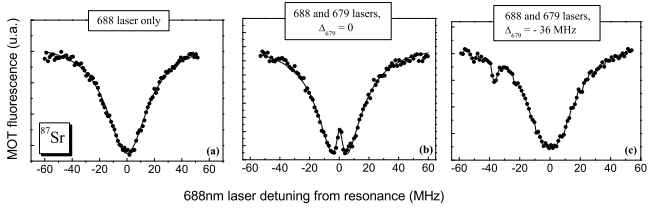}
\end{center}
\caption{\small{MOT fluorescence signals as a function of the
688\,nm laser detuning from the $^3P_1,F=9/2-\,^3S_1,F'=11/2$
transition of $^{87}$Sr. (a) 688\,nm laser only; (b) and (c) with
an additional 679\,nm laser at fixed frequency, detuned by
$\Delta_{679}$ from $^3P_0,F=9/2-\,^3S_1,F'=11/2$ resonance. The
linewidth of the CPT resonance is 3\,MHz due to power and Zeeman
broadening. Measurements recorded with laser intensities
$I_{688}$=1.3\,mW.cm$^{-2}$ and $I_{679}$=130\,mW.cm$^{-2}$.
}}\label{scanCPTRaman}
\end{figure}

In Fig.\,\ref{scanCPTRaman} is plotted the MOT fluorescence as the
688\,nm laser is swept around the $^3P_1,F=9/2-\,^3S_1,F'=11/2$
transition of $^{87}$Sr. For the signal shown in
Fig.\,\ref{scanCPTRaman}.a the 688\,nm laser alone is sent to the
atoms. As depicted in the previous section, a decrease in the MOT
fluorescence is observed as the 688\,nm laser pumps atoms from
$^3P_1$ to $^3P_0$ and $^3P_2$. In Fig.\,\ref{scanCPTRaman}.b and
\ref{scanCPTRaman}.c, the 679\,nm laser also shines the trap and
the CPT resonance is observed. When the 679\,nm laser is tuned
exactly onto the $^3P_0,F=9/2-\,^3S_1,F'=11/2$ transition, the CPT
resonance is detected as an increase of the MOT fluorescence at
the center of the broad 688\,nm resonance
(Fig.\,\ref{scanCPTRaman}.b). This increase is due to the atoms
projected onto the dark state $|\psi_{NC}\rangle$ which decay to
the ground state and recover the trapping process. If we gradually
increase the 679\,nm laser detuning, the CPT resonance moves
sidewards and reverses sign when it reaches the wings of the
688\,nm resonance (Fig.\,\ref{scanCPTRaman}.c). We then detect
atoms projected onto $|\psi_{C}\rangle$ which are optically pumped
to $^3P_2$, a process which increases the MOT losses as compared
to the situation with the 688\,nm laser alone.

The CPT resonance allows for a direct measurement of the
$^3P_0,F=9/2-\,^3P_1, F'=9/2$ fine structure splitting of
$^{87}$Sr. Three measurements have been performed with each of the
$^3S_1$ hyperfine levels as a third state in the $\Lambda$ system.
For $^{88}$Sr and others $^3P_1$ hyperfine states of $^{87}$Sr,
the CPT resonances are blurred due to the higher sensitivity of
the involved states to magnetic field. Only a deformation of the
688\,nm resonance has been observed which did not allow for
accurate measurements.

\begin{figure}
\begin{center}
\includegraphics[width=\columnwidth]{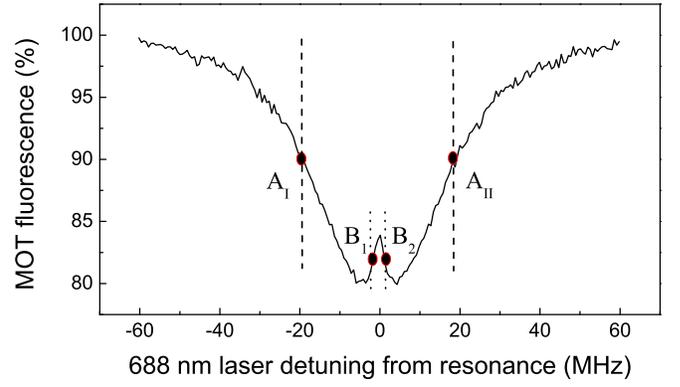}
\end{center}
\caption{\small{688\,nm and 679\,nm lasers are both locked by
frequency modulation of the 688\,nm laser. From fluorescence
measurements at frequencies $A_I$ et $A_{II}$ the 688\,nm laser is
locked to $^3P_1-\,^3S_1$ resonance. Measurements at $B_1$ et
$B_{2}$ are used to locked the 679\,nm to $^3P_0-\,^3S_1$.
Experiment performed tuning the 688\,nm laser around the $^3P_1,
F=9/2-\,^3S_1,F'=11/2$ resonance of
$^{87}$Sr.}}\label{fig:asservissementCPT}
\end{figure}

The fine structure measurements are performed with the frequency
of both 688\,nm and 679\,nm lasers locked to the atomic signal.
Three digital servo loops run in parallel. Servo $A$ locks the
688\,nm laser frequency to the $^3P_1-\,^3S_1$ transition. This is
done by alternating measurements at frequencies $A_I$ and $A_{II}$
indicated in Fig. \ref{fig:asservissementCPT}. With servo $B$, the
688\,nm laser is locked onto the CPT peak thanks to measurements
at frequencies $B_1$ and $B_{2}$ (Fig.
\ref{fig:asservissementCPT}). Finally servo $C$ controls the
frequency of the 679\,nm laser so as to equalize the frequencies
of servos $A$ and $B$. Measurements for servo $A$ and $B$ are
interlaced with variable durations. The fine structure measurement
is deduced from the difference of the average frequencies of
servos $A$ and $C$. A frequency resolution of $10\,$kHz is
achieved in a few minutes of averaging.

\begin{figure}
\begin{center}
\includegraphics[width=\columnwidth]{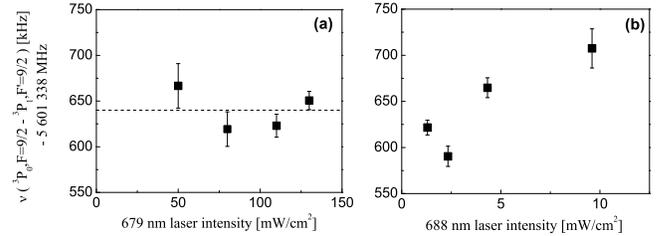}
\end{center}
\caption{\small{Frequency measurements of
$^3P_0,F=9/2-\,^3P_1,F=9/2$ splitting as a function of (a) the
679\,nm laser intensity and (b) the 688\,nm laser intensity. Both
lasers are locked to resonance: the 679\,nm laser to $^3P_0,
F=9/2-\,^3S_1,F'=11/2$ and the 688\,nm laser to $^3P_1,
F=9/2-\,^3S_1,F'=11/2$.}}\label{fig:CPTLightShift}
\end{figure}

With this scheme, both lasers are locked on their respective
atomic resonance which enables light shift free measurements.
Indeed, no shift is observed as a function of the 679\,nm laser
intensity (see Fig.\,\ref{fig:CPTLightShift}.a). When the 688\,nm
laser intensity is increased however, we do observe a shift
(Fig.\,\ref{fig:CPTLightShift}.b). As previously reported for the
$^3P_1-\,^3S_1$ measurements (Fig.\,\ref{fig:effetP688}), this
effect is attributed to the Zeeman effect induced by the MOT field
gradient. Increasing the laser intensity enhances the number of
atoms excited away from the zero magnetic field region. It is to
be noted that using a 679\,nm laser at high intensity ($\sim$
100\,mW.cm$^{-2}$) does not enhance the Zeeman effect because the
688\,nm laser performs the spatial selection.

Again the Zeeman effect is the dominant systematic frequency shift
in this experiment. It has been shown in the previous section that
it is minimized using lasers in the particular polarization
{\sf{Polar1}} and addressing only atoms close to the center of the
MOT. The sensitivity of the CPT resonance to the Zeeman effect is
5 to 6 times smaller than for the 688\,nm measurements of
$^{87}$Sr. We attribute a 50\,kHz uncertainty to the
$^3P_0,F=9/2-\,^3P_1, F'=9/2$ fine structure measurements. They
have been performed via the three hyperfine levels of $^3S_1$. We
finally obtain:
\[
\nu\,(^3P_0, F=9/2-\,^3P_1,F'=9/2)=5\,601\,338\,670\,(50)\,\mathrm{kHz}
\]

\begin{table}
\begin{center}
\begin{tabular}{c|c|l|c}
\hline\hline
\multicolumn{2}{c|}{$^3P_0-\,^3S_1$} &  \multicolumn{2}{c}{frequency (kHz)}\\
  \hline
$^{88}$Sr &J=0 -J'=1& 441 332 751 300 (700)&(a)\\
\hline
 & F=9/2-F'=7/2 & 441 335 740 420 (350)&(b)\\
\cline{2-4}
$^{87}$Sr& F=9/2-F'=9/2 & 441 333 301 370 (350)&(b)\\
\cline{2-4}
& F=9/2-F'=11/2 & 441 330 320 270 (350)&(b)\\
\hline\hline
\end{tabular}
\end{center}
\caption{\small{ $^3P_0-\,^3S_1$ atomic frequencies. (a) : Direct
measurement. (b) : Computed from CPT measurements and values from
Table \ref{tab:freq688}.}}\label{tab:freq679}
\end{table}

From this measurement and the values of table\,\ref{tab:freq688}
the frequency of all the $^3P_0, F=9/2-\,^3S_1, F'$ transitions
are deduced (Table\,\ref{tab:freq679}). The isotope shift of the
$^3P_0$ state can also be extracted:
\[
\Delta_{87,88}\,[^3P_0]=62.9 \;(1.3)\;\mathrm{MHz}
\]

\begin{table}
\begin{center}
\begin{tabular}{ l|rcr}
\hline  \hline
\multicolumn{4}{c}{$\Delta_{87,88}$ (kHz)}\\
\hline
$^3P_0$   & &62\,900 &(1\,300) \\
$^3P_1$   & &62\,150 &(70) \\
$^3S_1$   & &54\,900 &(300) \\
\hline  \hline
\end{tabular}
\end{center}
\caption{\small{Isotope shifts between $^{87}$Sr and $^{88}$Sr
(referenced to the ground state).}}\label{tab:decalageIso}
\end{table}

\begin{table}
\begin{center}
\begin{tabular}{c|c|l}
  \hline \hline
&Rubbmark {\it et al} & \multicolumn{1}{c}{This work}\\
\raisebox{1.5ex}[0cm][0cm]{$^{88}$Sr }&$\nu$(MHz) &  \multicolumn{1}{c}{$\nu$(MHz)}  \\
 \hline
$^1S_0-\,^3P_1$ &434\,829\,300&434 829 121.30 (0.02)\\
 \hline
 $^3P_1-\,^3S_1$&435\,731\,500&435 731 697.2 (0.5)\\
 \hline
  $^3P_0-\,^3S_1$&441\,332\,600&441 332 751.3 (0.7)\\
 \hline\hline
\end{tabular}
\end{center}
\caption{\small{Comparison between our frequency measurements and
values computed from Rubbmark {\it et al} \cite{Rubbmark78}.
}}\label{tab:comparaison}
\end{table}

We have determined all the frequencies of the $^1S_0-\,^3P_1$,
$^3P_1-\,^3S_1$ and $^3P_0-\,^3S_1$ transitions for $^{87}$Sr and
$^{88}$Sr. From these measurements the $^3S_1$ hyperfine constants
and the $^3P_0$, $^3P_1$ and $^3S_1$ isotope shifts could be
extracted (Table\,\ref{tab:decalageIso}). Frequency measurements
performed for $^{88}$Sr can be compared to the values of
Ref.\,\cite{Rubbmark78}. Authors of this reference estimate their
accuracy to 0.002\,nm ($\simeq$1.2\,GHz) but notice that their
values are in agreement to better than 0.0005\,nm (300\,MHz) with
those reported in Ref.\,\cite{Sullivan38}. As shown in
Table\,\ref{tab:comparaison}, it appears they are wrong by
100\,MHz to 200\,MHz only.

One of the goals of the above experiments was to yield an accurate
knowledge of the frequency of $^{1}S_0-^3P_0, F=9/2$ clock
transition of $^{87}$Sr. We find $\nu=429\,228\,004\,340
(70)$\,kHz. This indirect estimate was a decisive step toward the
direct observation of this transition which we describe in the
following section with more details than in
\cite{Courtillot03PRA}.

\section{Direct measurement of the $^{1}S_0-\,^3P_0$ transition}

For $^{87}$Sr, the $^{1}S_0-\,^3P_0$ transition at 698\,nm is
slightly allowed by hyperfine coupling of $^3P_0$ to $^1P_1$ and
$^3P_1$. Its theoretical linewidth is estimated to
1\,mHz\,\cite{KatoPal03,Porsev04}. Detection of such a narrow line
is a technical challenge with atoms at 2\,mK even if its frequency
is already known. To obtain a detectable signal we have developed
an original technique based on the MOT dynamics and the use of a
698\,nm laser frequency sweep. The principle consists in inducing
in the MOT a leak to $^3P_0$ with a 698\,nm laser. Detection of a
small number of atoms in this state is made difficult by the
absence of cycling transition from this state. The detection of
the $^{1}S_0-\,^3P_0$ transition is then performed by measuring
the loss induced in the MOT with the 461\,nm fluorescence of the
trapped atoms.

In order to excite the largest possible fraction of atoms, the
698\,nm laser is used at the highest available power, 14\,mW. The
laser beam is sent four times through the MOT forming two standing
waves. Both waves are tilted by 5\,mrad with respect to each-other
and by 45$^\circ$ with respect to vertical. Their waist radius is
1.3\,mm. With these parameters, the resonance is broadened by
saturation to 1.8\,kHz. On the other hand, the Doppler broadening
for our 2\,mK atomic sample is 1.5\,MHz (full width at
half-maximum (FWHM)). Only one atom out of $10^{3}$ is then
expected to be resonant with the laser at a given time.

However, the MOT dynamics can lead to an amplification of the
transfer rate to $^3P_0$. Indeed, with the laser parameters given
above, the duration of a $\pi$ pulse on the narrow transition is
0.5\,ms. This is about 100 times shorter than the MOT lifetime. It
is then in principle possible to enhance the fraction of excited
atoms by the same factor if one accumulates atoms in $^3P_0$. The
induced MOT loss should then reach several \%. This works only if
two conditions are fulfilled. First, the transfer rate to $^3P_0$
has to remain constant. This is at first sight not the case since
the 698\,nm laser excitation creates a dip in the velocity
distribution of atoms in the ground state. Second, atoms once in
the $^3P_0$ state have to actually escape the trapping process. If
they stay resonant with the 698\,nm laser frequency, they flip
back to the ground state by stimulated emission. To ensure both
conditions, several methods can be thought of. For instance, the
MOT beams could be used to rethermalize atoms and fill the dip in
the the ground state velocity distribution. For the escape
condition, optical pumping to $^3P_2$ state could be performed
with a laser tuned to the $^{3}P_0-\,^3S_1$ resonance as in
Ref.\,\cite{Takamoto03}. Alternatively, both conditions can be
simultaneously fulfilled by sweeping the laser frequency. This is
doable if the frequency ramp is perfectly controlled so as to not
degrade the accuracy of the frequency measurements.

We have used a simpler method based on the Doppler effect induced
by the acceleration of atoms by gravity. The experiment is
operated sequentially. By means of acousto-optic modulators we
alternate capture and cooling phases with the 461\,nm lasers and
probe phases with the 698\,nm laser. During probe phases atoms are
freely falling. Due to the $45^\circ$ angle formed by the 698\,nm
beam and the vertical, the frequency sweep caused by gravity
amounts to 10\,kHz/ms. Both atoms transferred to the $^3P_0$ state
and the corresponding dip in the velocity distribution of the
ground state are then detuned from the laser excitation. The
pulsed operation of the experiment also prevents any light shift
of the ground state by the MOT beams during probe phases.

We have developed a simple model of the loss induced in the trap
by excitation to $^3P_0$. We compute the transition probability
for an atom probed by two counterpropagating beams of 28\,mW each,
hence averaging down the effect of the interference due to the
5\,mrad tilt between the actual beams. We also neglect spontaneous
emission and treat the atom as a two level system. For an atom of
initial velocity $\vec{v}_0$ we numerically solve the optical
Bloch equations with a time dependent detuning from resonance
$\Delta(t)$. Accounting for the gravity acceleration $\vec{g}$ it
writes down as:
\begin{eqnarray}
\label{eq:delta}&&\Delta(t) = \Delta_0 +\varepsilon\;\alpha \,t \\
&&\mathrm{with:} \quad \Delta_0= \delta +\varepsilon\; \vec{k} .\vec{v}_0\\
\label{eq:doppler}&& \qquad \quad \; \alpha = \frac{1}{\sqrt{2}}
\;\vec{k}. \vec{g} \simeq 2\pi\times 10\,\mathrm{kHz/ms}.
\end{eqnarray}
Here $\delta$ is the laser detuning in the laboratory frame,
$\vec{k}$ the probe laser wave vector and $\varepsilon = \pm 1$
depending on the direction of the laser beam interacting with the
atoms. We have neglected interferences between counterpropagating
laser beams considering they address different atomic velocity
classes. This is true for large enough detuning, typically
$|\delta|>10\,\Omega$, with $\Omega$ the Rabi frequency.
Accounting for both beams would lead to a sub-Doppler structure at
the center of the resonance which was not resolved experimentally.

\begin{figure}
\begin{center}
\includegraphics[width=\columnwidth]{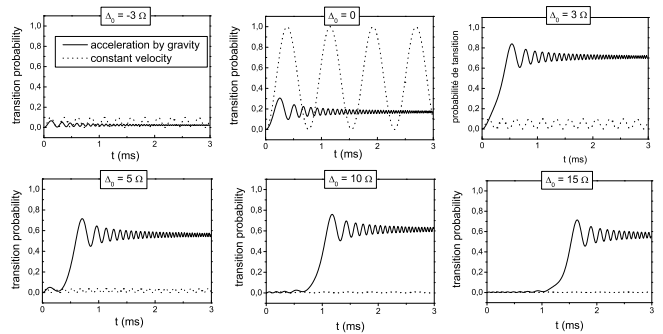}
\end{center}
\caption{\small{Transition probabilities as a function of the
interaction time for various detunings $\Delta_0$ at $t=0$ (with
$\Omega/2\pi=920$\,Hz). Continuous lines: atoms accelerated by
gravity. Dashed lines: atoms having a constant
velocity.}}\label{fig:proba1Atome698}
\end{figure}

The transition probabilities obtained for various values of
$\Delta_0$ in the case $\varepsilon=\,-1$ are plotted in solid
lines in Fig.\,\ref{fig:proba1Atome698}. When atoms with a blue
initial detuning ($\Delta_0>0$) are brought onto resonance by the
acceleration of gravity, the transition probability rapidly
increases and then tends to stabilize around a high value, for
instance 0.56 when $\Delta_0=5\,\Omega$. Also shown on
Fig.\,\ref{fig:proba1Atome698} (dashed lines) is the transition
probability which would be obtained in the absence of gravity,
{\it i.e.} for a constant value $\Delta(t)=\Delta_0$. It is clear
that if the interaction time is sufficient, a significant
transition probability is obtained for a much broader range of
initial atomic velocity when gravity is accounted for.

The evolution of the transition probability of an initially blue
detuned atom is remindful of adiabatic rapid passage. Indeed our
experimental parameters are at the edge of the adiabatic transfer
condition\,\cite{Messiah64ANG}:
\begin{equation}
\dot{\Delta} \ll 2\;\Omega^2.
\end{equation}
Our experimental parameters yield:
\[
\frac{\dot{\Delta}}{2\;\Omega^2}\simeq 0.94.
\]

\begin{figure}
\begin{center}
\includegraphics[width=\columnwidth]{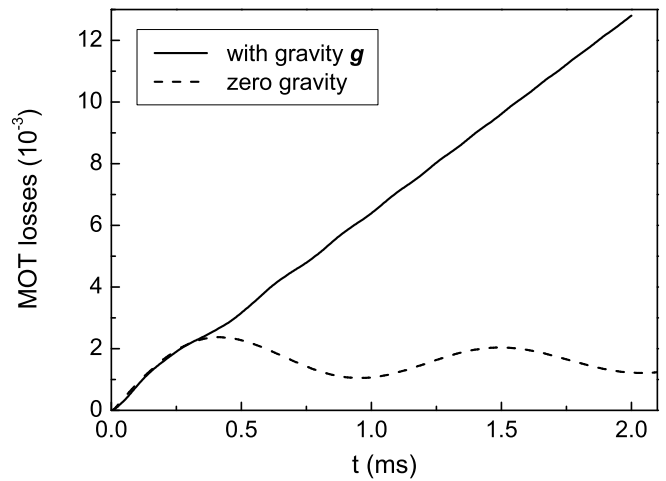}
\end{center}
\caption{\small{Relative loss induced in the MOT as a function of
the interaction time in the absence (dashed line) and in presence
(solid line) of gravity.}}\label{fig:probaTotale698}
\end{figure}

The loss induced in the MOT per probe pulse is given by the above
transition probability after integration over the whole atomic
velocity distribution. In Fig.\,\ref{fig:probaTotale698} are
plotted the results obtained in the presence (solid line) and in
the absence (dashed line) of gravity. At zero gravity the loss
rapidly stabilizes around a small value: $1.6\times 10^{-3}$ of
the total number of trapped atoms. This value is close to the
intuitive value given by the ratio of saturation over Doppler
broadenings. By contrast, gravity enables to induce loss in the
MOT that keep increasing with the interaction time: more and more
atoms are brought onto resonance due to gravity. An even faster
frequency sweep would lead to a higher increase of the loss in the
MOT. For instance, with a 100\,kHz/ms ramp, a further increase of
40\% is expected. This would however be at the cost of an
experimental complication. The frequency ramp should be perfectly
controlled and synchronized with the experimental time sequence.

Our time sequence consists in alternating phases of cooling and
capture of 3\,ms duration with probe phases of 1\,ms duration.
This sequence, which was experimentally optimized, results from a
trade-off between the MOT capture efficiency, the ballistic
expansion of the atomic cloud during probe phases, and the
efficiency of excitation to $^3P_0$. From the results of our model
and given the 40\,ms lifetime of the MOT with this time sequence
parameters, we expect a contrast of the resonance of 6\%.

\begin{figure}
\begin{center}
\includegraphics[width=\columnwidth]{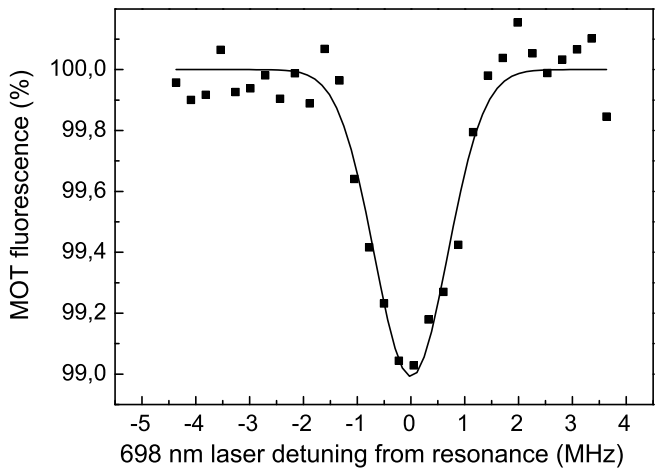}
\end{center}
\caption{Direct observation of $^{1}S_0-\,^3P_0$ transition of $^{87}$Sr. The line is broadened by the Doppler
effect to 1.4\,MHz (FWHM) due to the temperature of the atoms. The maximal value of the MOT fluorescence
corresponds to $3\times 10^6$ trapped atoms.
}\label{fig:698resonance} 
\end{figure}

In Fig.\,\ref{fig:698resonance} is shown the observed
$^{1}S_0-\,^3P_0$ resonance. The contrast is 1\%, 6 times lower
than the expected value. The discrepancy between experimental and
expected value may have several causes. Neglecting the angle
between both standing waves may lead to an overestimate of the MOT
loss by a factor of nearly 2. In addition, the ballistic expansion
of the atomic cloud has been neglected in the model. During probe
phases, atoms are moving away from the 698\,nm beam center due to
their residual velocity. The actual Rabi frequency thus decreases
with time. This also leads to a decrease of the transfer rate to
$^3P_0$.

We have locked the 698\,nm laser to the $^{1}S_0-\,^{3}P_0$
resonance with a digital servo loop. The error signal is derived
from MOT fluorescence measurements performed sequentially on both
sides of the resonance. The duration of each fluorescence
measurement is a multiple of the time sequence period used to
probe the atoms. The error signal is integrated and fed back to
the synthesizer driving the offset-phase lock of the probe laser
ECLD2 (Fig\,\ref{fig:exp}). The mean frequency of the measurements
is 429\,228\,004\,235 kHz with a residual statistical uncertainty
of 15\,kHz.

At this level, systematic effects are negligible. The residual
Doppler effect is well below 1\,kHz due to the standing wave
configuration. The Zeeman effect amounts to less than 1\,kHz in
spite of the MOT field gradient since Land{\'e} factors of
$^{1}S_0$ and $^{3}P_0$ states are extremely small
(Tab.\,\ref{tab:annexeLande}). Finally measurements performed with
imperfect extinction by the acousto-optic modulators show that the
residual light shift by the 461\,nm light is also below 1\,kHz.
The Doppler profile being shifted from the atomic resonance by the
recoil frequency\,\cite{Demtroder}, the measured mean frequency
has to be corrected by 4.7\,kHz\footnote{The value published
in\,\cite{Courtillot03PRA} was erroneously not corrected for the
recoil.}. We finally obtain:
\[
\nu\,(^1S_0-\,^3P_0)= 429\,228\,004\,230\;(15)\;\mathrm{kHz.}
\]

This value is in agreement with the indirect determination, the
difference between both frequencies being 1.4\,$\sigma$.

\section*{Conclusion}

We have measured the frequency of the $^1S_0-\,^3P_1$
(tab.\,\ref{tab:freq689}), $^3P_1-\,^3S_1$
(tab.\,\ref{tab:freq688}) and $^3P_0-\,^3S_1$
(tab.\,\ref{tab:freq679}) optical transitions for both strontium
isotopes $^{87}$Sr et $^{88}$Sr. The reported accuracy is in the
100\,kHz range, an improvement by three orders of magnitude as
compared to previous values\,\cite{Rubbmark78}. This was made
possible by the versatility of femto-second laser based optical
frequency measurements and in turn nicely illustrates the
potential of this technique. It is worth noting that with
traditional harmonic frequency chain measurements, four quite
different (and complex) experiments should have been built, one
for each specific wavelength: 679\,nm, 688\,nm, 689\,nm and
finally 698\,nm.

These measurements lead to a better knowledge of atomic Sr
yielding accurate values of the hyperfine structure parameters of
state $^3S_1$ and of the isotope shift of states $^3P_0$, $^3P_1$
and $^3S_1$. Thanks to the indirect determination of the energy of
state $^3P_0$ the $^1S_0-\,^3P_0$ clock transition of $^{87}$Sr
could be directly detected. Although this transition is extremely
forbidden (its natural linewidth is 1\,mHz) we have been able to
observe it on a sample at a temperature of 2\,mK by making use of
the magneto-optical trap dynamics. We have induced a leak in the
MOT to state $^3P_0$ which enhanced the contrast of the transition
by one order of magnitude as compared to simple saturation
broadening. We performed a direct measurement of the clock
transition frequency which is found in agreement with the indirect
determination. The technique used here could be applied to detect
other strongly forbidden lines like the slightly broader
$^1S_0-\,^3P_0$ transition of fermionic Yb. The lifetime of Yb
atoms in a MOT using the $^1S_0-\,^3P_1$
transition\,\cite{Kuwamoto99} is typically two orders of magnitude
longer than our Sr MOT and the contrast of the resonance could
then approach 100\,\%.

Although this clock transition is interesting in itself, whatever
the scheme used to probe the atoms, the frequency measurement of
the $^1S_0-\,^3P_0$ of $^{87}$Sr is the first step towards the
construction of a new generation of atomic clocks: optical clocks
with trapped atoms\,\cite{KatoPal03,Takamoto03}. It seems
realistic with these clocks to reach in a close future a
fractional frequency instability around
$10^{-16}\,\tau^{-1/2}$\,\cite{Quessada03} together with an
accuracy in the $10^{-17}$ range\,\cite{KatoPal03}.

We thank Ouali Acef, S{\'e}bastien Bize, Andr\'{e} Clairon, Michel
Lours, Giorgio Santarelli and J-J. Zondy for helpful discussions,
and the optoelectronic group of the university of Bath (UK) for
providing the photonic cristal fiber of the frequency chain. A. B.
acknowledges his grant from the european Research Training Network
CAUAC. BNM-SYRTE is Unit\'e Associ\'ee au CNRS (UMR 8630).

\section*{Annexe: Land\'{e} factors\label{ann:lande}}

\begin{table}[h]
\begin{tabular}{c|c|c||c|c||c|c|c|c}
\hline \hline

 &\multicolumn{2}{c||}{$^1S_0$} &\multicolumn{2}{c||}{$^3P_0$} & \multicolumn{4}{c}{$^1P_1$ }\\
\cline{2-9}
 &$^{87}$Sr&$^{88}$Sr &$^{87}$Sr&$^{88}$Sr&\multicolumn{3}{c|}{ $^{87}$Sr}  &  $^{88}$Sr \\
\hline
  F& 9/2&0&9/2&0&7/2 & 9/2& 11/2 & 1\\
\hline
  $g_F$ &$-1.3\times10^{-4}$ &0&$-\,6\times10^{-5}$ &0&-2/9&4/99&2/11&1\\
\hline\hline
\end{tabular}

\vspace{\baselineskip}

\begin{tabular}{c|c|c|c|c||c|c|c|c}
\hline\hline
& \multicolumn{4}{c||}{$^3P_1$ }& \multicolumn{4}{c}{$^3S_1$}\\
\cline{2-9}
& \multicolumn{3}{c|}{ $^{87}$Sr} &   $^{88}$Sr & \multicolumn{3}{c|}{ $^{87}$Sr}  &  $^{88}$Sr \\
\hline
  F&7/2 & 9/2& 11/2 & 1 &7/2 & 9/2& 11/2 & 1\\
\hline
  $g_F$ & -1/3&2/33 & 3/11& 3/2 & -4/9&8/99 &4/11&2\\
\hline\hline
\end{tabular}
\caption{\small{Land\'{e} factors.}}\label{tab:annexeLande}
\end{table}

\pagebreak

\end{document}